\documentclass[12pt]{article}
\usepackage{graphics}

\def\<~{\stackrel{<}{\mbox{\scriptsize $\sim$}}}
\def\>~{\stackrel{>}{\mbox{\scriptsize $\sim$}}}
\def\eq#1{(\ref{#1})}

\begin{document}

\title{
Braneworld Quintessential Inflation and Sum of Exponentials Potentials}
\author{
Carl L. Gardner\\
{\em gardner@math.asu.edu}\\
Department of Mathematics and Statistics\\ 
Arizona State University\\
Tempe AZ 85287-1804
}
\date{}

\maketitle
\thispagestyle{empty}

\begin{abstract}

  Quintessential inflation---in which a single scalar field plays the
  role of the inflaton {\em and}\/ quintessence---from a sum of
  exponentials potential $V = A \left(e^{5 \varphi} + e^{\sqrt{2}
      \varphi}\right)$ or $V = A \left(e^{5 \varphi} +
    e^\varphi\right)$ or a cosh potential $V = 2A \cosh(5 \varphi)$ is
  considered in the context of five-dimensional gravitation with
  standard model particles confined to our 3-brane.  Reheating is
  accomplished via gravitational particle production and the universe
  undergoes a transition from primordial inflation to radiation
  domination well before big bang nucleosynthesis.  The transition to
  an accelerating universe due to quintessence occurs near $z \approx
  1$, as in $\Lambda$CDM.

  Braneworld quintessential inflation can occur for potentials with or
  without a minimum, and with or without eternal acceleration and an
  event horizon.  The low $z$ behavior of the equation of state
  parameter $w_\phi$ provides a clear observable signal distinguishing
  quintessence from a cosmological constant.

\end{abstract}

\section{Introduction}

In five-dimensional gravitation with standard model particles confined
to our 3-brane, the Friedmann equation is
modified~\cite{BDL1}--\cite{Shiro} at high energies: the square $H^2$
of the Hubble parameter acquires a term quadratic in the energy
density, allowing~\cite{Maartens} slow-roll inflation to occur for
potentials that would be too steep to support inflation in the
standard Friedmann-Robertson-Walker cosmology.  In this model,
quintessential inflation---in which a single scalar field plays the
role of the inflaton {\em and}\/ quintessence---can occur for a sum of
exponentials or cosh potential, a type of potential that arises
naturally in M/string theory for combinations of moduli fields or
combinations of the dilaton and moduli fields.

Inflation ends in this model as the quadratic term in energy density
in $H^2$ decays to roughly the same order of magnitude as the standard
linear term (technically, when the slow-roll parameter $\epsilon$ =
1).  Reheating can be accomplished~\cite{Copeland} via gravitational
particle production at the end of inflation.  The universe
subsequently~\cite{Copeland} undergoes a transition from the era
dominated by the scalar field potential energy to an era of
``kination'' dominated by the scalar field kinetic energy, and then to
the standard radiation dominated era well before big-bang
nucleosynthesis (BBN).  With a sum of exponentials or cosh potential,
the universe evolves at this point according to the quintessence/cold
dark matter (QCDM) model.  The transition to an accelerating universe
due to quintessence occurs late in the matter dominated era near $z
\approx 1$, as in $\Lambda$CDM.

We will assume a flat universe after inflation.  In the QCDM model,
the total energy density $\rho = \rho_m + \rho_r + \rho_\phi =
\rho_c$, where $\rho_c$ is the critical energy density for a flat
universe and $\rho_m$, $\rho_r$, and $\rho_\phi$ are the energy
densities in (nonrelativistic) matter, radiation, and the
quintessential inflation scalar field $\phi$, respectively.  Ratios of
energy densities to the critical energy density will be denoted by
$\Omega_m = \rho_m/\rho_c$, $\Omega_r = \rho_r/\rho_c$, and
$\Omega_\phi = \rho_\phi/\rho_c$, while ratios of present energy
densities $\rho_{m0}$, $\rho_{r0}$, and $\rho_{\phi0}$ to the present
critical energy density $\rho_{c0}$ will be denoted by $\Omega_{m0}$,
$\Omega_{r0}$, and $\Omega_{\phi0}$, respectively.

Using WMAP3~\cite{WMAP3} central values, we will set $\Omega_{\phi0}$
= 0.74, $\Omega_{r0} = 8.04 \times 10^{-5}$, $\Omega_{m0} = 1 -
\Omega_{\phi0} - \Omega_{r0} \approx 0.26$, and $\rho_{c0}^{1/4} =
2.52 \times 10^{-3}$ eV, with the present time $t_0 = 13.7$ Gyr after
the big bang.

For the sum of exponentials potential, we will take a monotonically
increasing function of $\varphi$ ($\varphi$ will evolve from
$\varphi_i \gg 1$ at the end of inflation to $|\varphi_0| \sim 1$
today)
\begin{equation}
	V(\varphi) = A \left( e^{\lambda \varphi} + B e^{\mu \varphi} \right)
\label{V}
\end{equation}
where $A$ and $B$ are positive constants, $\lambda > \mu > 0$,
$\varphi = \phi/M_P$, the (reduced) Planck mass $M_P = 2.44 \times
10^{18}$ GeV, and with $\lambda = 5$, $\mu = \sqrt{2}$ or 1, and $B =
1$ for the simulations.  We will also discuss the simple exponential
potential $V(\varphi) = A e^{\lambda \varphi}$ considered in
Ref.~\cite{Copeland} and the cosh potential $V(\varphi) = 2 A
\cosh(\lambda \varphi)$ with $\lambda = 5$.  For all four potentials,
$V \sim A e^{\lambda \varphi}$ for $\varphi \gg 1$ (which will be the
case during inflation and gravitational particle production).  The
constant $A \sim \rho_{c0}$ for quintessential inflation (see
Table~1).

Note that the BBN ($z \sim 10^9$--$10^{11}$), cosmic microwave
background (CMB) ($z \sim 10^3$--$10^5$), and large-scale structure
(LSS) ($z \sim 10$--$10^4$) bounds $\Omega_\phi \<~ 0.1$ are satisfied
in all the simulations below (Figs.~\ref{fig-cosh-Omega},
\ref{fig-sum2-Omega}, and \ref{fig-sum-Omega}) and that the transition
from the era of scalar field dominance to the radiation era occurs
around $z \approx 10^{15}/\lambda$.

\begin{figure}[htbp]
\center{
\scalebox{1.1}{\includegraphics{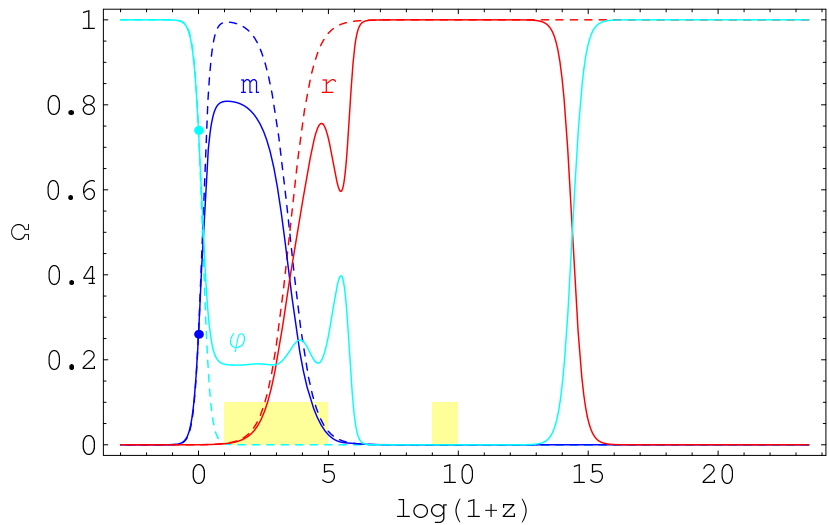}}
}
\caption{$\Omega$ for $V = A\left(e^{4 \varphi} + e^{0.1 \varphi}\right)$
  (solid) vs.\ $\Lambda$CDM (dotted).  The light yellow rectangles are
  the bounds $\Omega_\phi \le 0.1$ from LSS, CMB, and BBN.}
\label{fig-sum4-Omega}
\end{figure}

Quintessential inflation in the standard cosmology from a sum of
exponentials potential was considered in Ref.~\cite{Barreiro:1999zs}
for $\lambda = 20$ and $\mu = 0.5$ or $-20$.
Ref.~\cite{Majumdar:2001mm} extended the analysis to the braneworld
case where $H^2$ has a quadratic term in energy density, for $\lambda
\approx 20$ and $\mu \approx - \lambda$.  Braneworld quintessential
inflation is also discussed in Ref.~\cite{Nunes:2002wz}, for $\lambda
= 4$ and $\mu = 0.1$.  Only one simulation for the sum of exponentials
potential braneworld scenario is presented, in
Ref.~\cite{Nunes:2002wz} for the equation of state parameter $w_\phi$.
We find that the case $\lambda = 4$ and $\mu = 0.1$ violates even the
looser CMB bound $\Omega_\phi \le 0.2$ (see
Fig.~\ref{fig-sum4-Omega}), and that the recent average of the scalar
field equation of state parameter $\overline{w}_0 = -0.68$ is too
large.

The current investigation furthermore presents detailed simulations
and analyses of the evolution of the scalar field and its equation of
state, the fractional energy densities in the scalar field, radiation,
and matter, and the acceleration parameter in braneworld
quintessential inflation for the sum of exponentials and cosh
potentials.

\section{Cosmological Equations}

The homogeneous scalar field---since it is confined to the
brane---still obeys the Klein-Gordon equation
\begin{equation}
	\ddot{\phi} + 3 H \dot{\phi} = -\frac{d V}{d \phi} \equiv - V_\phi ~.
\label{phi}
\end{equation}
The Hubble parameter $H$ is related to the scale factor $a$ and the
energy densities in matter, radiation, and the scalar field
through the braneworld modified~\cite{Maartens} Friedmann equation
\begin{equation}
	H^2 = \left( \frac{\dot{a}}{a} \right)^2 = \frac{\rho}{3 M_P^2}
        \left( 1 + \frac{\rho}{2 \sigma} \right)
        + \Lambda_4 + \frac{{\cal E}}{a^4}
\label{H-original}
\end{equation}
where  the energy density
\begin{equation}
	\rho = \rho_\phi + \rho_m + \rho_r ,~~
        \rho_\phi = \frac{1}{2} \dot{\phi}^2 + V(\phi) ,
\label{rho}
\end{equation}
$\sigma$ is the four-dimensional brane tension, $\Lambda_4$ is the
four-dimensional cosmological constant, and ${\cal E}$ is a constant
embodying the effects of bulk gravitons on the brane.  We will set the
four-dimensional cosmological constant to zero.  The ``dark
radiation'' term ${{\cal E}/a^4}$ can be ignored here since it will
rapidly go to zero during inflation.  Thus for our purposes the
modified Friedmann equation takes the form
\begin{equation}
	H^2 = \frac{\rho}{3 M_P^2} \left( 1 + \frac{\rho}{2 \sigma} \right) .
\label{H}
\end{equation}
In the low-energy limit $\rho \ll \sigma$, the Friedmann equation
reduces to its standard form $H^2 = \rho/(3 M_P^2)$.

The conservation of energy equation for matter, radiation, and the
scalar field is
\begin{equation}
	\dot{\rho} + 3 H (\rho + P) = 0
\label{energy}
\end{equation}
where $P$ is the pressure.  Except near particle-antiparticle
thresholds, $P_m$ = 0 and $P_r = \rho_r/3$.  Equation~\eq{energy}
gives the evolution of $\rho_m$ and $\rho_r$, and the Klein-Gordon
equation~\eq{phi} for the weakly coupled scalar field, with
\begin{equation}
	P_\phi = \frac{1}{2} \dot{\phi}^2 - V(\phi) .
\end{equation}

The acceleration equation
\begin{equation}
        \frac{\ddot{a}}{a} = - \frac{1}{6 M_P^2} \left(
        \rho + 3 P + \frac{\rho}{\sigma} \left( 2 \rho + 3 P \right)
        \right)
\end{equation}
follows from Eqs.~\eq{H} and \eq{energy}.  While inflation occurs in
the low-energy (standard cosmology) limit when $w \equiv P/\rho < -
1/3$, for inflation to occur in the high-energy limit $w < - 2/3$.

We will use the logarithmic time variable $\tau = \ln(a/a_0) =
-\ln(1+z)$.  Note that for de Sitter space $\tau = H_\Lambda t$, where
$H_\Lambda^2 = \rho_\Lambda/(3 M_P^2)$, and that $H_\Lambda t$ is a
natural time variable for the era of $\Lambda$-matter domination (see
e.g.\ Ref.~\cite{alpha}).  For $0 \le z \le z_{BBN} \sim 10^{10}$,
$-23.03 \le \tau \le 0$, and for $0 \le z \le z_{Pl} \sim 2.8 \times
10^{31}$, $-72.41 \le \tau \le 0$.

In Eq.~\eq{rho}, we will make the simple approximations
\begin{equation}
        \rho_r = \rho_{r0} e^{-4 \tau} ,~~
	\rho_m = \rho_{m0} e^{-3 \tau} .
\end{equation}

For the spatially homogeneous scalar field $\phi$, the equation
of state parameter $w_\phi = w_\phi(z) = P_\phi/\rho_\phi$.
The recent average of $w_\phi$ is defined as
\begin{equation}
	\overline{w}_{0} = \frac{1}{\tau} \int_0^\tau w_\phi d \tau .
\label{wbar}
\end{equation}
We will take the upper limit of integration $\tau$ to correspond to
$z$ = 1.75.  The SNe Ia observations~\cite{Riess:2004nr} bound the
recent average $\overline{w}_0 < -0.76$ (95\% CL) assuming
$\overline{w}_0 \ge -1$, and measure the transition redshift $z_t =
0.46 \pm 0.13$ from deceleration to acceleration (it is probably
premature at this point to say more than that $z_t \approx 1$).

For numerical simulations, the cosmological equations should be put
into a scaled, dimensionless form.  Equations~\eq{phi} and~\eq{H} can
be cast~\cite{quint} in the form of a system of two first-order
equations in $\tau$ plus a scaled version of $H$:
\begin{equation}
	\tilde{H} \varphi' = \psi
\label{phi-tilde-1}
\end{equation}
\begin{equation}
	\tilde{H} (\psi' + \psi) = - 3 \tilde{V}_\varphi
\label{phi-tilde-2}
\end{equation}
\begin{equation}
	\tilde{H}^2 = \tilde{\rho} 
        \left( 1 + \frac{\tilde{\rho}}{2 \tilde{\sigma}} \right)
\label{scaled-H}
\end{equation}
\begin{equation}
	\tilde{\rho} = \frac{1}{6} \psi^2 + \tilde{V} + 
	\tilde{\rho}_m + \tilde{\rho}_r
\label{scaled-rho}
\end{equation}
where $\psi \equiv e^{2\tau} \dot{\varphi}/H_0$, $\tilde{H} =
e^{2\tau} H/H_0$, $\tilde{V} = e^{4 \tau} V/\rho_{c0}$,
$\tilde{V}_\varphi = e^{4 \tau} V_\varphi/\rho_{c0}$, $\tilde{\rho} =
e^{4\tau} \rho/\rho_{c0}$, $\tilde{\rho}_m = e^{4\tau}
\rho_m/\rho_{c0} = \Omega_{m0} e^\tau$, $\tilde{\rho}_r = e^{4 \tau}
\rho_r/\rho_{c0} = \Omega_{r0}$, $\tilde{\sigma} = e^{4\tau}
\sigma/\rho_{c0}$, and where a prime denotes differentiation with
respect to $\tau$: $\varphi' = d\varphi/d\tau$, etc.

This scaling results in a set of equations that
is numerically more robust, especially before the time of BBN.

\begin{figure}[htbp]
\center{
\scalebox{1.1}{\includegraphics{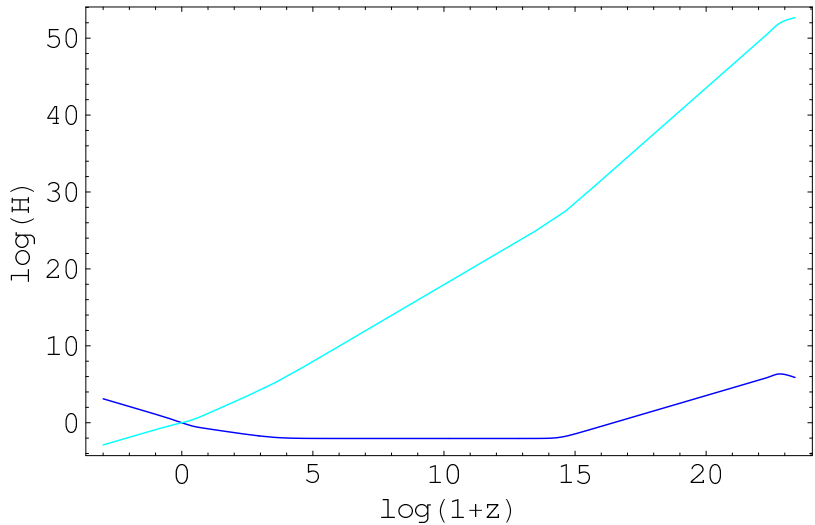}}
}
\caption{Log$_{10}$ of $\tilde{H}$ (blue, bottom) and $H/H_0$ (cyan,
  top) vs.\ $\log_{10}(1+z)$ for $V = A\left(e^{5 \varphi} +
    e^{\sqrt{2} \varphi}\right)$.}
\label{fig-H}
\end{figure}

Figure~\ref{fig-H} illustrates that while $\tilde{H}$ spans only ten
orders of magnitude between $z_i \sim 10^{23}$ and the present,
$H/H_0$ spans more than fifty orders of magnitude.

\section{Evolution of the Braneworld Universe}

In this section, we will closely follow the analyses of Refs.~\cite{Copeland}
and \cite{Maartens}.

\subsection{Slow-Roll Braneworld Inflation}

The inflationary slow-roll parameter $\epsilon$ is given by~\cite{Maartens}
\begin{equation}
        \epsilon \equiv - \frac{\dot{H}}{H^2} \approx
        \frac{M_P^2}{2} \left( \frac{V_\phi}{V} \right)^2 
        \frac{1+V/\sigma}{(1+V/(2 \sigma))^2} ~.
\label{epsilon}
\end{equation}
(The slow-roll parameter 
\begin{equation}
        \eta = M_P^2 \frac{V_{\phi \phi}}{V}
        \frac{1}{1+V/(2 \sigma)} \approx \epsilon
\end{equation}
for the potentials considered here.)  Inflation occurs for $\epsilon <
1$.  The slow-roll parameter can be approximated during inflation by
\begin{equation}
        \epsilon \approx \frac{2 \lambda^2 \sigma}{V}
\end{equation}
for the sum of exponentials potential, the simple exponential
potential, and the cosh potential, since $V \gg \sigma$ and $\varphi
\gg 1$ for our models during inflation.  Inflation ends when $\epsilon
= 1$, implying~\cite{Copeland} that the potential $V_e$ at the end of
inflation is
\begin{equation}
        V_e \approx 2 \lambda^2 \sigma .
\label{Ve}
\end{equation}
To evaluate $V_e$ and $\sigma$, the COBE-measured amplitude of
primordial density perturbations is matched against the theoretical
value at $N$ = 50 e-folds from the end of inflation, where
\begin{equation}
        N = \int_{t_N}^{t_e} H dt
        \approx \frac{1}{M_P^2} \int_{\phi_e}^{\phi_N} \frac{V}{V_\phi}
        \left( 1 + \frac{V}{2 \sigma} \right) d\phi \approx 
        \frac{1}{2 \lambda^2 \sigma} (V_N - V_e)
\end{equation}
yielding $V_N \approx (N+1) V_e$.
Now matching against the amplitude~\cite{Maartens} of density fluctuations
\begin{equation}
        \left( 2 \times 10^{-5} \right)^2 =
        A_S^2 \approx \frac{1}{75 \pi^2 M_P^6} \frac{V^3}{V_\phi^2}
        \left( 1 + \frac{V}{2 \sigma} \right)^3 \approx
        \frac{1}{600 \pi^2} \frac{V^4}{M_P^4 \lambda^2 \sigma^3}
\end{equation}
gives~\cite{Copeland}
\begin{equation}
        V_e^{1/4} \approx 1.11 \times 10^{15}~{\rm GeV}/\lambda ,~~
        \sigma^{1/4} \approx 9.37 \times 10^{14}~{\rm GeV}/\lambda^{3/2} .
\end{equation}

\subsection{Gravitational Reheating and Initial Conditions}

At the end of inflation, gravitational particle
production~\cite{Ford}--\cite{Ferreira:1997hj} results in a radiation
energy density~\cite{Copeland}
\begin{equation}
        \frac{\rho_r}{\rho_\phi} \approx 0.01 g_p \frac{H_e^4}{V_e}
        \approx 2.8 \times 10^{-4} g_p \frac{V_e^3}{M_P^4 \sigma^2}
        \approx 4.86 \times 10^{-17} g_p
\end{equation}
where $g_p \approx$ 10--100 is the number of particle species that are
gravitationally produced.  This relation fixes the ``initial
conditions'' redshift $z_i \approx 3.89 \times 10^{23}
g_p^{1/4}/\lambda$ at which gravitational particle production occurs,
from the scaling
\begin{equation}
        \rho_r \approx 4.86 \times 10^{-17} g_p \rho_\phi \approx
        g_p \left( \frac{10^{11}~{\rm GeV}}{\lambda} \right)^4
        \approx (1 + z_i)^4 \rho_{r0}
\end{equation}
corresponding~\cite{Copeland} to a reheating temperature
\begin{equation}
        T_e \sim \left( \frac{g_p}{g_*} \right)^{1/4}
        \frac{10^{11}~{\rm GeV}}{\lambda}
\end{equation}
where $g_*(T)$ is the effective number of massless degrees of freedom
at temperature $T$.  The universe undergoes~\cite{Copeland} a
transition from the era dominated by the scalar field potential energy
to an era of kination ($\rho_\phi \sim 1/a^6$) dominated by the scalar
field kinetic energy as the $\rho^2/(2 \sigma)$ term in $H^2$ becomes
small compared with $\rho$.  For this era of kination to occur, the
potential $V$ must be sufficiently steep.  Since during kination
$\rho_r/\rho_\phi \sim a^2$, the universe eventually makes a
transition~\cite{Copeland} to the standard radiation dominated era,
around a temperature $T \approx 10^3$ GeV$/\lambda$, well before
big-bang nucleosynthesis at $T \approx$ 1 MeV.  For the sum of
exponentials or cosh potential, the universe then evolves according to
the QCDM model.  The transition to an accelerating universe due to
quintessence occurs late in the matter dominated era near $z \approx
1$, as in $\Lambda$CDM.

The initial conditions at the end of inflation will be set in the
high-energy, slow-roll limit.  The initial value for $\phi_i$ is
specified by Eq.~\eq{Ve}.  The initial value for $\dot{\phi}_i$
follows from the high-energy, slow-roll limits
\begin{equation}
	H^2 \approx \frac{\rho^2}{6 M_P^2 \sigma} \approx
        \frac{V^2}{6 M_P^2 \sigma}
\end{equation}
and
\begin{equation}
	\ddot{\phi} + 3 H \dot{\phi} \approx
        \sqrt{\frac{3}{2}} \frac{V}{M_P \sqrt{\sigma}} \dot{\phi}
        \approx - V_\phi \approx - \lambda \frac{V}{M_P}
\end{equation}
or~\cite{Maartens}
\begin{equation}
	\frac{1}{2} \dot{\phi}_i^2 \approx \frac{V_e}{6} \approx
        \frac{1}{3} \lambda^2 \sigma .
\label{phidot}
\end{equation}

\section{Simulations}

For the computations below, we will use
Eqs.~\eq{phi-tilde-1}--\eq{scaled-rho} with initial conditions
specified at $z_i$ by $\varphi_i$ and $\dot{\varphi}_i \propto \psi_i$
(Eqs.~\eq{Ve} and \eq{phidot}).  We set $\lambda$ = 5 and $g_p$ = 100.
The constant $A$ in the potentials is adjusted so that
$\Omega_{\phi0}$ = 0.74.  This involves the usual single fine tuning.
The final time $t_f$ is set by $z_f = - 0.999$ corresponding to $t_f
\sim$ 100 Gyr.

First we briefly summarize the properties of the exponential
potential, and then turn our attention to the quintessential inflation
potentials.

\begin{table}[ht]
\center{
\begin{tabular}{|l c c c c|} \hline 
$V(\varphi)/A$ & $A/\rho_{c0}$ & $\varphi_i$ & $\varphi_0$ & 
        $\overline{w}_0$ \\ \hline
$2 \cosh(5 \varphi)$ & 0.4 & 48.0 & $-0.03$ & $-0.87$ \\
$e^{5 \varphi} + e^{\sqrt{2} \varphi}$ & 5.7 & 47.4 & $-1.60$ & $-0.78$ \\
$e^{5 \varphi} + e^\varphi$ & 2.2 & 47.6 & $-1.18$ & $-0.89$ \\
\hline
\end{tabular}
}
\caption{Simulation results for the cosh and sum of exponentials potentials.}
\label{potentials}
\end{table}

\subsection{Exponential Potential}

The exponential potential $V(\varphi) = A e^{\lambda
  \varphi}$~\cite{Ferreira:1997hj}--\cite{Doran:2002ec} can be derived
from M-theory~\cite{Townsend:2001ea} or from $N$ = 2, 4D gauged
supergravity~\cite{Andrianopoli:1996cm}.

For the exponential potential with $\lambda^2 > 3$, the cosmological
equations have a global attractor with $\Omega_\phi = 3/\lambda^2$
during the matter dominated era (during which $w_\phi = 0$); and with
$\lambda^2 > 4$, the cosmological equations have a global attractor
with $\Omega_\phi = 4/\lambda^2$ during the radiation dominated era
(during which $w_\phi = 1/3$).  For $\lambda^2 < 3$, the cosmological
equations have a late time attractor with $\Omega_\phi = 1$ and
$w_\phi = \lambda^2/3 - 1$.


For $\lambda = \sqrt{2}$ and $\rho_m = 0$, $\ddot{a} \rightarrow 0$
asymptotically; if $\rho_m > 0$, the universe eventually enters a
future epoch of deceleration.  In either case, there is no event
horizon.  For $\lambda < \sqrt{2}$, the universe enters a period of
eternal acceleration with an event horizon.  For $\lambda > \sqrt{2}$,
the universe eventually decelerates and there is no event horizon.

The $\Lambda$CDM cosmology is approached for $\lambda \le 1/\sqrt{3}$.
Significant acceleration occurs only for $\lambda \<~ \sqrt{3}$.  For
$\lambda = \sqrt{3}$, $\overline{w}_0$ is much too high; for a viable
present-day QCDM cosmology, $\lambda \le \sqrt{2}$~\cite{quint} in the
exponential potential.

\subsection{Cosh Potential}


In the cosh potential model 
\begin{equation}
        V = 2A \cosh(5 \varphi) \approx \rho_\Lambda \cosh(5 \varphi) ,
\end{equation}
dark energy derives from the value of the potential near its minimum.
This is the simplest way to ``correct'' the exponential potential to
incorporate quintessence.

\begin{figure}[htbp]
\center{
\scalebox{1.1}{\includegraphics{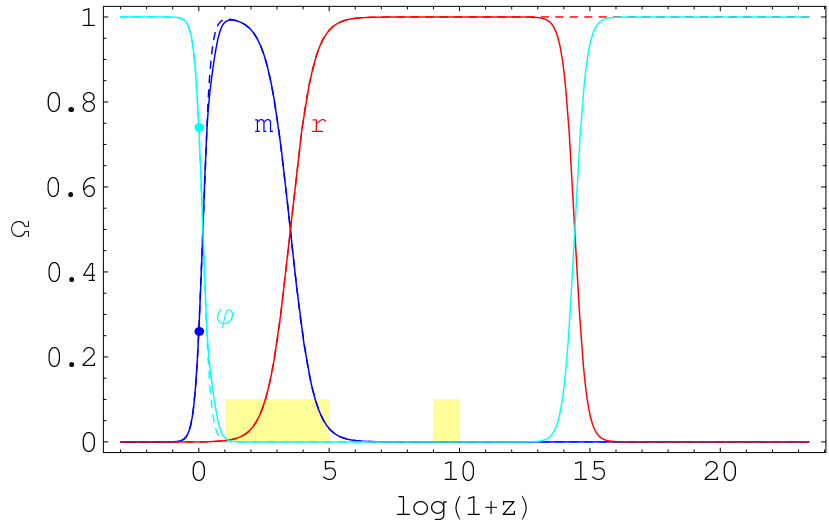}}
}
\caption{$\Omega$ for $V = 2A \cosh(5 \varphi)$ (solid)
  vs.\ $\Lambda$CDM (dotted).}
\label{fig-cosh-Omega}
\end{figure}

\begin{figure}[htbp]
\center{
\scalebox{1.1}{\includegraphics{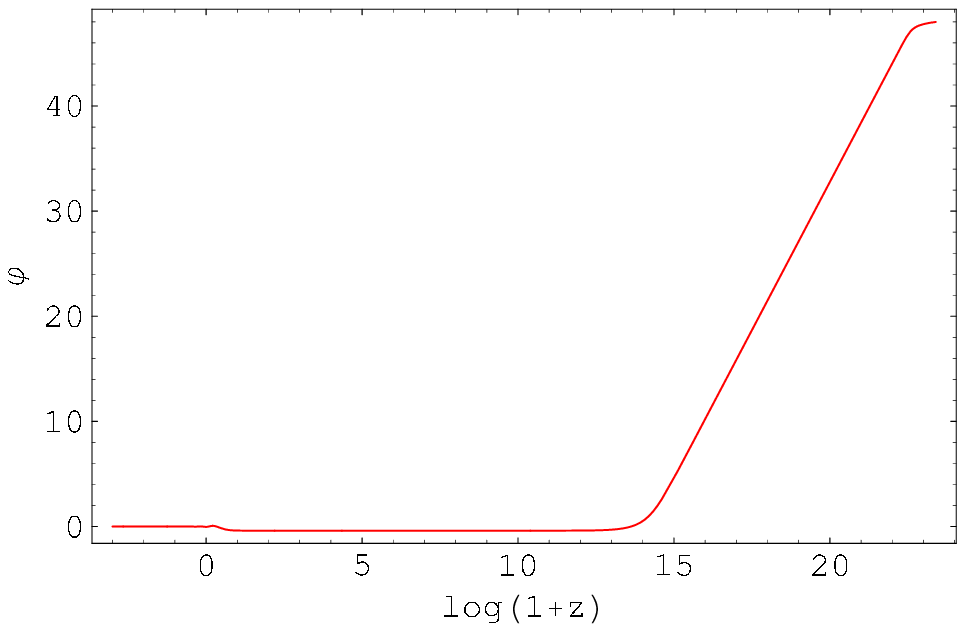}}
}
\caption{$\varphi$ for $V = 2A \cosh(5 \varphi)$.}
\label{fig-cosh-phi}
\end{figure}

\begin{figure}[htbp]
\center{
\scalebox{1.1}{\includegraphics{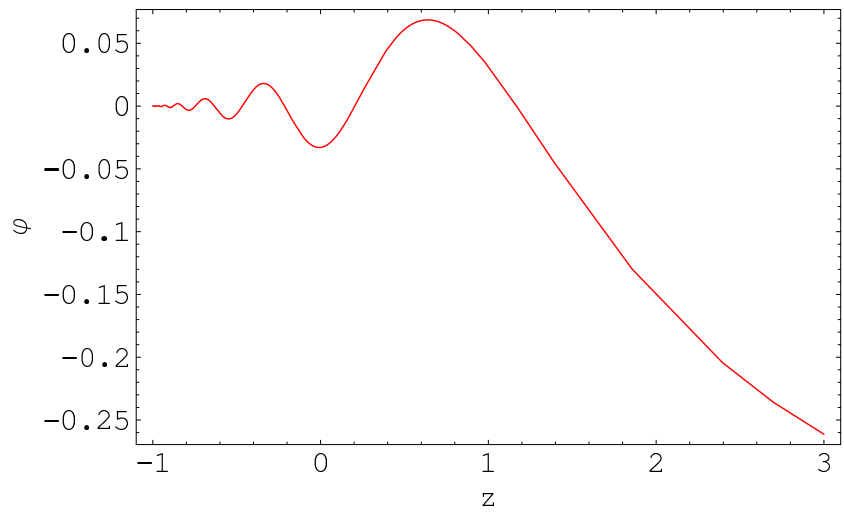}}
}
\caption{Oscillating behavior of $\varphi$ near $z = 0$ for $V = 2A
  \cosh(5 \varphi)$.}
\label{fig-cosh-phi0}
\end{figure}

\begin{figure}[htbp]
\center{
\scalebox{1.1}{\includegraphics{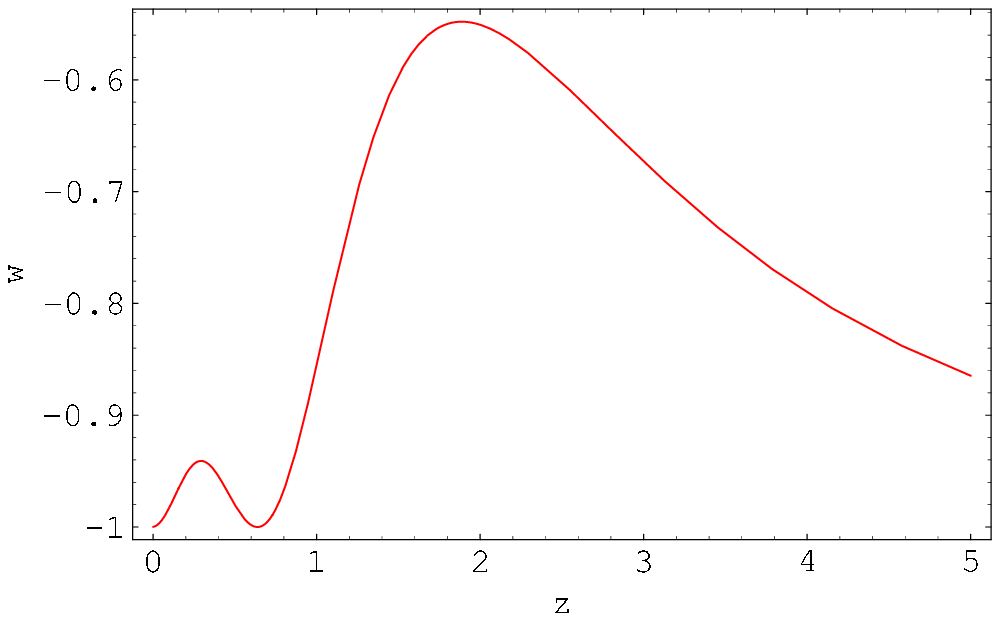}}
}
\caption{$w_\phi$ for $V = 2A \cosh(5 \varphi)$.}
\label{fig-cosh-w}
\end{figure}

\begin{figure}[htbp]
\center{
\scalebox{1.1}{\includegraphics{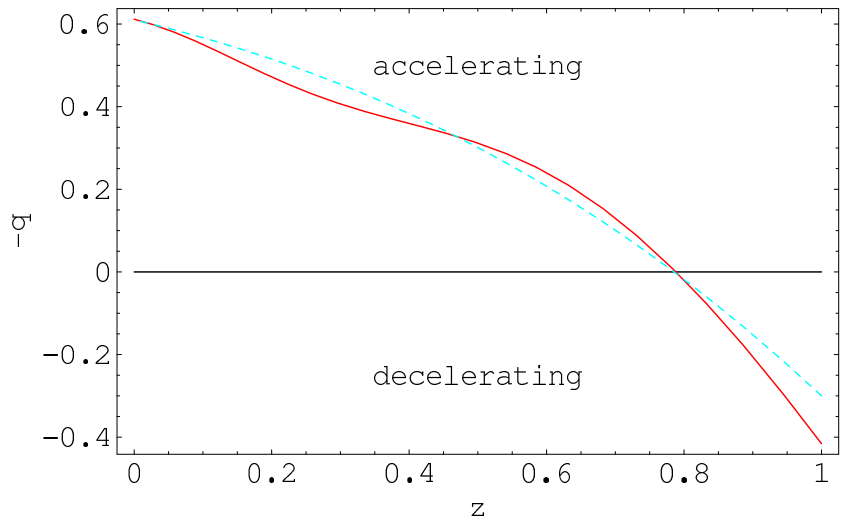}}
}
\caption{Acceleration parameter $-q$ for $V = 2A \cosh(5 \varphi)$
  (solid) vs.\ $\Lambda$CDM (dotted).}
\label{fig-cosh-acc}
\end{figure}

In the simulations for the cosh potential
(Figs.~\ref{fig-cosh-Omega}--\ref{fig-cosh-acc}), initially $\varphi
\approx 50$ and then evolves to $\varphi_0 = -0.03$ at $t_0$.  The
linear decay of $\varphi(\tau)$ in Fig.~\ref{fig-cosh-phi} during the
kination era occurs because $\rho_\phi \approx \frac{1}{2}
\dot{\phi}^2 \sim 1/a^6$ and thus $\varphi'$ is a (negative) constant
since both $\tilde{H}$ and $\psi$ are proportional to $e^{-\tau}$ in
Eq.~\eq{phi-tilde-1}.

Between $z \approx 2.5 \times 10^{14}$ (as the universe enters its
radiation dominated stage) and $z \approx 10$ (when $m_\phi^2 \approx
\lambda^2 V/M_P^2$ becomes of order $H^2 \approx \rho_m/(3 M_P^2)$),
$\varphi$ ``sits and waits'' at a small negative value.

With the cosh potential, $\varphi$ is very slowly oscillating and
decaying about $\varphi = 0$ at the minimum of the potential for $z
\<~ 1$ (we have neglected any present-day particle production by the
oscillating scalar field), with angular frequency $\omega \approx
\sqrt{3} \lambda H_\Lambda$ and decay time constant $\overline{t}
\approx 2 H_\Lambda^{-1}/3$ (see Fig.~\ref{fig-cosh-phi0}).

Note that $w_\phi \rightarrow -1$ for $t > t_0$
(Fig.~\ref{fig-cosh-w}).  The acceleration parameter $-q$ in
Fig.~\ref{fig-cosh-acc} closely mimics the $\Lambda$CDM curve.

\subsection{Sum of Exponentials Potential}

In the sum of exponentials potential~\eq{V}, a shift $\varphi
\rightarrow \varphi + \chi$ simply results in a redefinition of the
constants $A$ and $B$:
\begin{equation}
	V(\varphi) \rightarrow V(\varphi + \chi) = 
        {\cal A} \left( e^{\lambda \varphi} + 
        {\cal B} e^{\mu \varphi} \right) ,~~
        {\cal A} = e^{\lambda \chi} A ,~~
        {\cal B} = e^{(\mu - \lambda) \chi} B .
\end{equation}
In a realistic particle theory $A$ and $B$ would be set from first
principles; here we choose $B = 1$.  It will turn out then that $A
\sim \rho_{c 0}$ for quintessential inflation.

To produce primordial inflation in the braneworld scenario with the
sum of exponentials potential and the subsequent transition to the
standard radiation dominated universe satisfying the constraints on
$\Omega_\phi$, $\lambda \>~ 4.7$ (this agrees with the estimate
$\lambda \>~ 4.5$ given in Ref.~\cite{Copeland}); for present-day
quintessence, $\mu \le \sqrt{2}$~\cite{quint}.  We will choose
$\lambda = 5$ and $\mu = \sqrt{2}$ or 1 (these values appear naturally
in low-energy limits of M/string theory):
\begin{equation}
	V(\varphi) = A \left( e^{5 \varphi} +  e^{ \sqrt{2} \varphi} \right)
\end{equation}
or
\begin{equation}
	V(\varphi) = A \left( e^{5 \varphi} +  e^\varphi \right) .
\end{equation}

\begin{figure}[htbp]
\center{
\scalebox{1.1}{\includegraphics{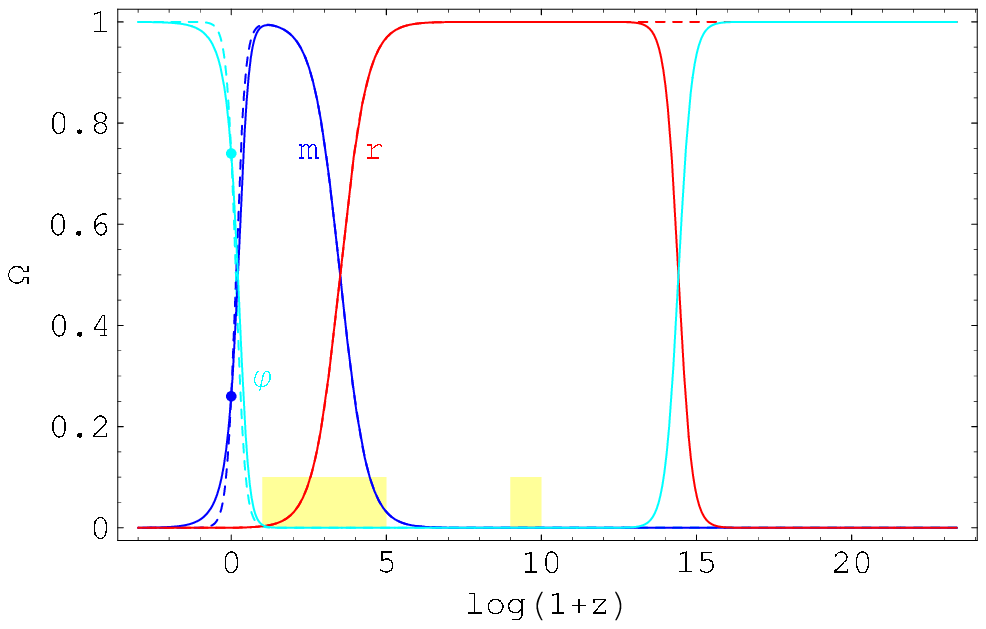}}
}
\caption{$\Omega$ for $V = A\left(e^{5 \varphi} + e^{\sqrt{2} \varphi}\right)$
  (solid) vs.\ $\Lambda$CDM (dotted).}
\label{fig-sum2-Omega}
\end{figure}

\begin{figure}[htbp]
\center{
\scalebox{1.1}{\includegraphics{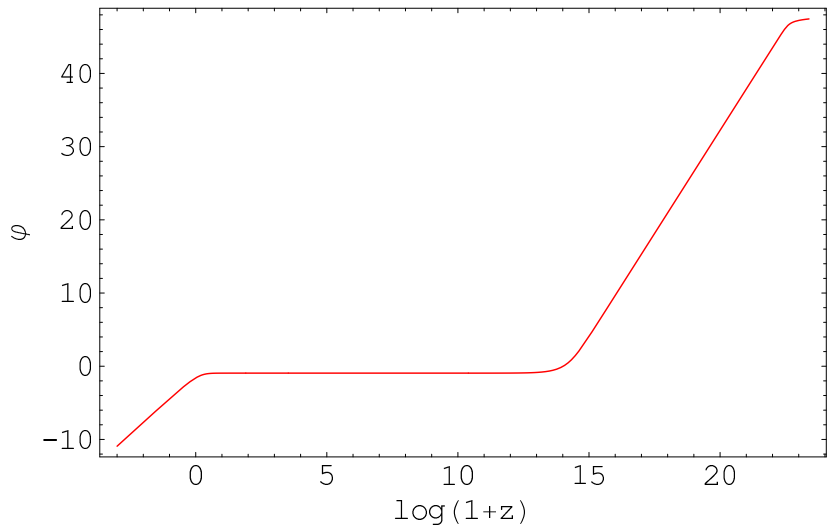}}
}
\caption{$\varphi$ for $V = A\left(e^{5 \varphi} + e^{\sqrt{2} \varphi}\right)$.}
\label{fig-sum2-phi}
\end{figure}

\begin{figure}[htbp]
\center{
\scalebox{1.1}{\includegraphics{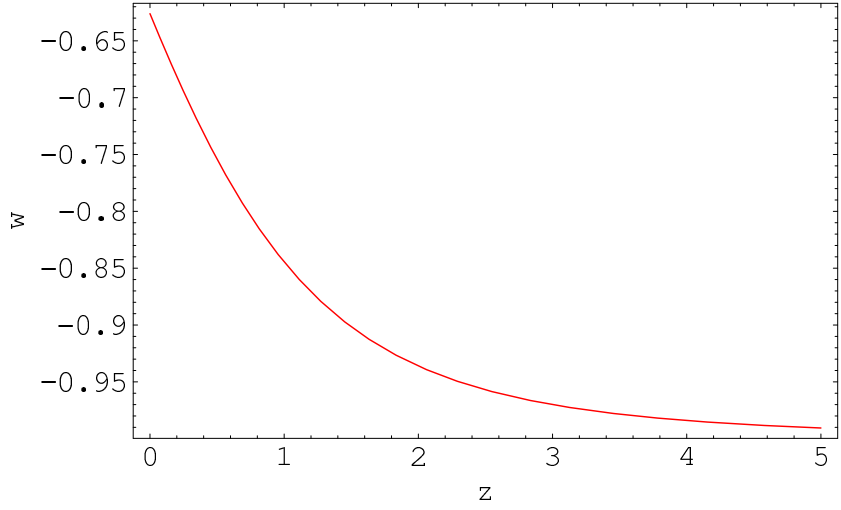}}
}
\caption{$w_\phi$ for $V = A\left(e^{5 \varphi} + e^{\sqrt{2} \varphi}\right)$.}
\label{fig-sum2-w}
\end{figure}

\begin{figure}[htbp]
\center{
\scalebox{1.1}{\includegraphics{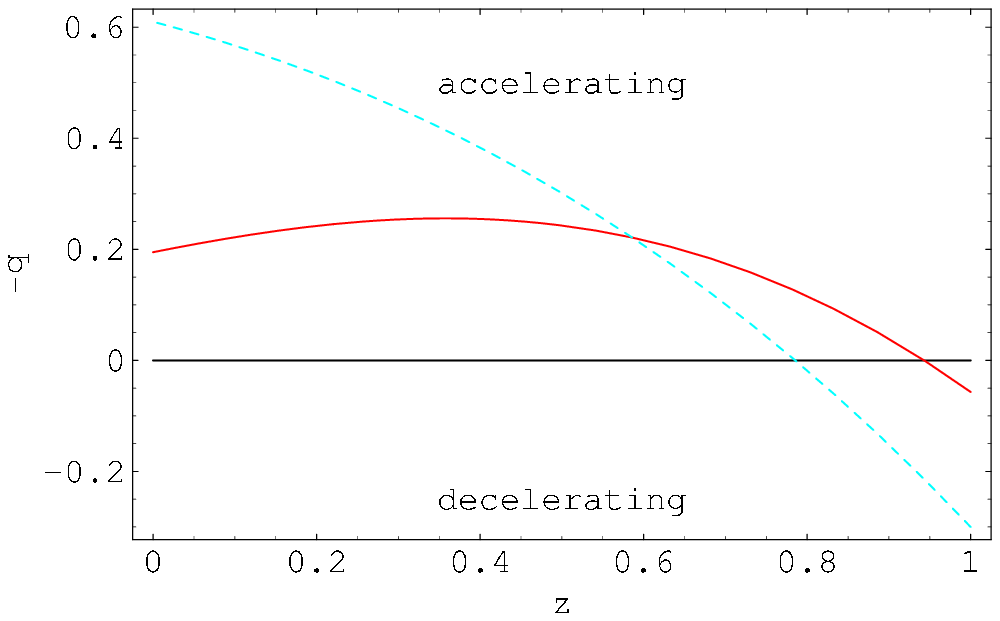}}
}
\caption{Acceleration parameter $-q$ for $V = A\left(e^{5 \varphi} +
  e^{\sqrt{2} \varphi}\right)$ (solid) vs.\ $\Lambda$CDM (dotted).}
\label{fig-sum2-acc}
\end{figure}

\begin{figure}[htbp]
\center{
\scalebox{1.1}{\includegraphics{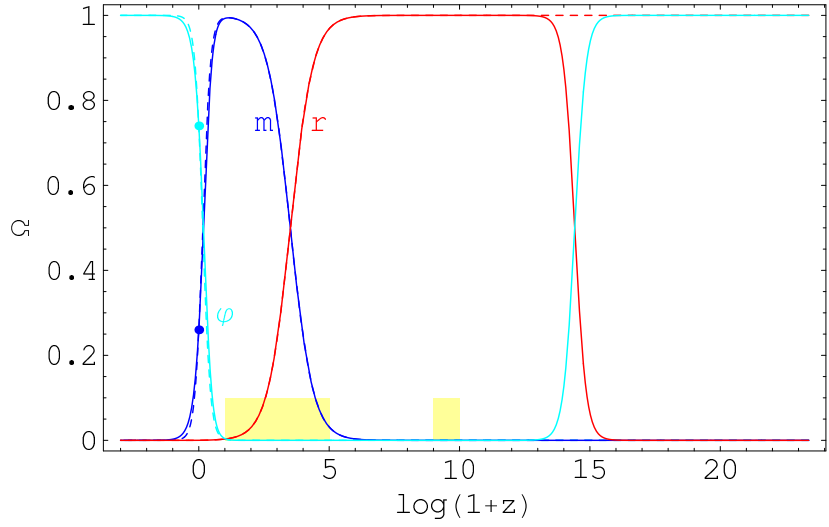}}
}
\caption{$\Omega$ for $V = A\left(e^{5 \varphi} + e^\varphi\right)$
  (solid) vs.\ $\Lambda$CDM (dotted).}
\label{fig-sum-Omega}
\end{figure}

\begin{figure}[htbp]
\center{
\scalebox{1.1}{\includegraphics{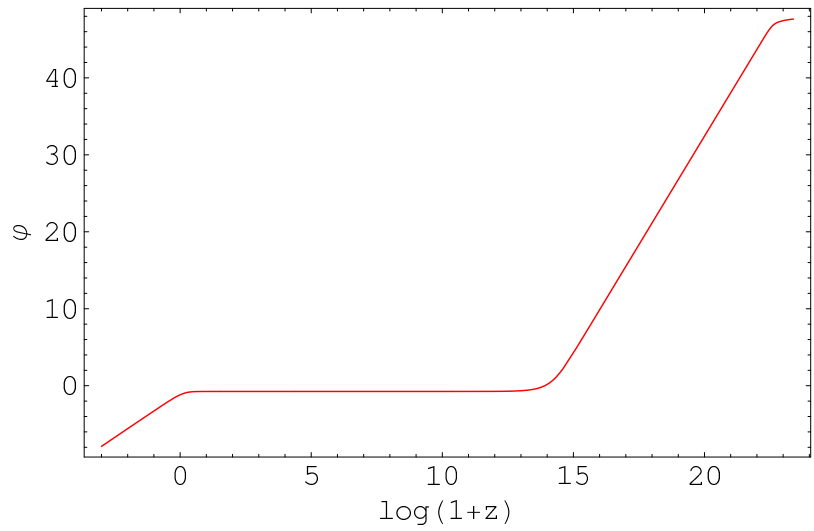}}
}
\caption{$\varphi$ for $V = A\left(e^{5 \varphi} + e^\varphi\right)$.}
\label{fig-sum-phi}
\end{figure}

\begin{figure}[htbp]
\center{
\scalebox{1.1}{\includegraphics{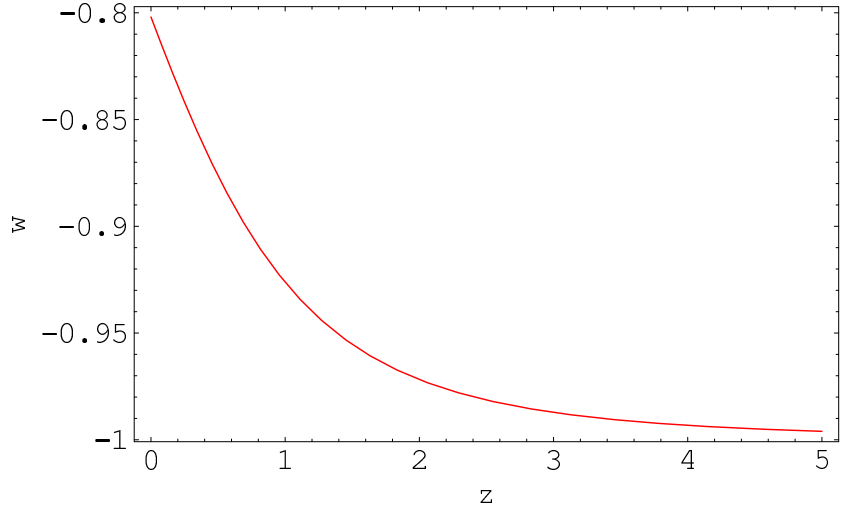}}
}
\caption{$w_\phi$ for $V = A\left(e^{5 \varphi} + e^\varphi\right)$.}
\label{fig-sum-w}
\end{figure}

\begin{figure}[htbp]
\center{
\scalebox{1.1}{\includegraphics{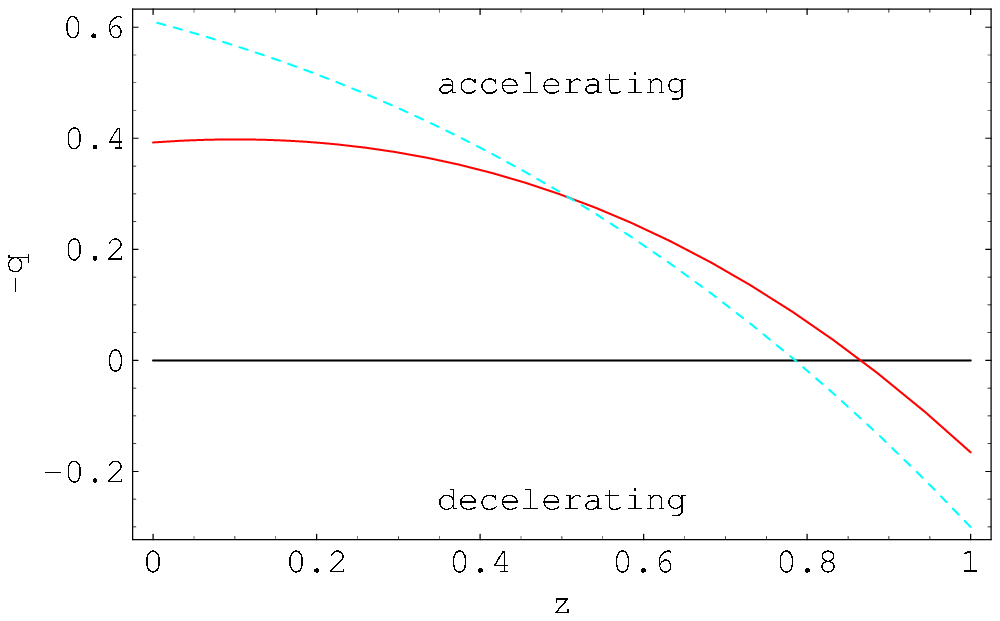}}
}
\caption{Acceleration parameter $-q$ for $V = A\left(e^{5 \varphi} +
  e^\varphi\right)$ (solid) vs.\ $\Lambda$CDM (dotted).}
\label{fig-sum-acc}
\end{figure}

For the sum of exponentials potentials
(Figs.~\ref{fig-sum2-Omega}--\ref{fig-sum-acc}), initially $\varphi
\approx 50$ and then evolves to $\varphi_0 \sim -1$.  Again note the
the linear decay of $\varphi(\tau)$ in Figs.~\ref{fig-sum2-phi} and
\ref{fig-sum-phi} during the kination era.

Between $z \approx 2.5 \times 10^{14}$ and $z \approx 2$ (when now
$m_\phi^2 \approx \mu^2 V/M_P^2$ becomes of order $H^2 \approx
\rho_m/(3 M_P^2)$), $\varphi$ ``sits and waits'' at a small negative
value, and then becomes more negative.  The linear decay of
$\varphi(\tau)$ after $t_0$ occurs because $\varphi$ is approaching
the late time attractor with $\Omega_\phi = 1$ and $w = \mu^2/3 - 1$
(Figs.~\ref{fig-sum2-w} and \ref{fig-sum-w}).  Whenever $w$ = const
and $H$ approximates its standard form, $\varphi(\tau)$ is linear in
$\tau$ since both $\tilde{H}$ and $\psi$ are proportional to
$e^{(1-3w)\tau/2}$ in Eq.~\eq{phi-tilde-1} ($w = 1$ during kination).

Note the differences in the acceleration parameter $-q$ in
Figs.~\ref{fig-sum2-acc} and \ref{fig-sum-acc} for the two cases $\mu
= \sqrt{2}$ (eventual deceleration with no event horizon) and $\mu =
1$ (eternal acceleration with an event horizon).

\section{Conclusion}

In the braneworld framework, the sum of exponentials and cosh
potentials yield natural quintessential inflation scenarios, motivated
by string/M theory.  The quintessential braneworld models also
provide---through an era of kination---a natural mechanism for the
transition of the universe from the primordial $\phi$-dominated era to
the era of radiation domination.

From the time of radiation domination to the distant future,
$\Omega_\phi$ mimics $\Omega_\Lambda$ for the quintessential inflation
potentials presented here.  In the $\Lambda$CDM model and in
braneworld quintessential inflation from a cosh or sum of exponentials
potential, there is a period between roughly 3.5 Gyr and 20 Gyr after
the big bang when $0.1 \le \Omega_\phi \le 0.9$.

Note that for the cosh and sum of exponentials potentials
$\overline{w}_0 \le -0.78$ (see Table~1), satisfying the observational
bound $\overline{w}_0 < -0.76$.

Braneworld quintessential inflation can occur for potentials with
(cosh) or without (sum of exponentials) a minimum, and with (sum of
exponentials with $\mu = 1$ or cosh) or without (sum of exponentials
with $\mu = \sqrt{2}$) eternal acceleration and an event horizon.

In both the cosh and sum of exponentials potentials considered here,
the low $z$ behavior of $w_\phi$ provides a clear observable signal
distinguishing quintessence from a cosmological constant.


\begin{thebibliography}{10}

\bibitem{BDL1}
  P.~Binetruy, C.~Deffayet, and D.~Langlois,
  Nucl.\ Phys.\ B {\bf 565}, 269 (2000)
  [arXiv:hep-th/9905012].

\bibitem{BDL2}
  P.~Binetruy, C.~Deffayet, U.~Ellwanger, and D.~Langlois,
  Phys.\ Lett.\ B {\bf 477}, 285 (2000)
  [arXiv:hep-th/9910219].

\bibitem{Shiro}
  T.~Shiromizu, K.~i.~Maeda, and M.~Sasaki,
  Phys.\ Rev.\ D {\bf 62}, 024012 (2000)
  [arXiv:gr-qc/9910076].

\bibitem{Maartens}
  R.~Maartens, D.~Wands, B.~A.~Bassett, and I.~Heard,
  Phys.\ Rev.\ D {\bf 62}, 041301 (2000)
  [arXiv:hep-ph/9912464].

\bibitem{Copeland}
  E.~J.~Copeland, A.~R.~Liddle, and J.~E.~Lidsey,
  Phys.\ Rev.\ D {\bf 64}, 023509 (2001)
  [arXiv:astro-ph/0006421].

\bibitem{WMAP3}
  D.~N.~Spergel {\it et al.},
  arXiv:astro-ph/0603449.

\bibitem{Barreiro:1999zs}
  T.~Barreiro, E.~J.~Copeland, and N.~J.~Nunes,
  Phys.\ Rev.\ D {\bf 61}, 127301 (2000)
  [arXiv:astro-ph/9910214].

\bibitem{Majumdar:2001mm}
  A.~S.~Majumdar,
  Phys.\ Rev.\ D {\bf 64}, 083503 (2001)
  [arXiv:astro-ph/0105518].

\bibitem{Nunes:2002wz}
  N.~J.~Nunes and E.~J.~Copeland,
  Phys.\ Rev.\ D {\bf 66}, 043524 (2002)
  [arXiv:astro-ph/0204115].

\bibitem{alpha}
  C.~L.~Gardner,
  Phys.\ Rev.\ D {\bf 68}, 043513 (2003)
  [arXiv:astro-ph/0305080].

\bibitem{Riess:2004nr}
  A.~G.~Riess {\it et al.}  [Supernova Search Team Collaboration],
  Astrophys.\ J.\  {\bf 607}, 665 (2004)
  [arXiv:astro-ph/0402512].

\bibitem{quint}
  C.~L.~Gardner,
  Nucl.\ Phys.\ B {\bf 707}, 278 (2005)
  [arXiv:astro-ph/0407604].

\bibitem{Ford}
  L.~H.~Ford,
  Phys.\ Rev.\ D {\bf 35}, 2955 (1987).

\bibitem{Spokoiny}
  B.~Spokoiny,
  Phys.\ Lett.\ B {\bf 315}, 40 (1993)
  [arXiv:gr-qc/9306008].

\bibitem{Ferreira:1997hj}
P.~G.~Ferreira and M.~Joyce,
Phys.\ Rev.\ D {\bf 58}, 023503 (1998)
[arXiv:astro-ph/9711102].

\bibitem{Wetterich:1987fm}
C.~Wetterich,
Nucl.\ Phys.\ B {\bf 302}, 668 (1988).

\bibitem{Copeland:1997et}
E.~J.~Copeland, A.~R.~Liddle, and D.~Wands,
Phys.\ Rev.\ D {\bf 57}, 4686 (1998)
[arXiv:gr-qc/9711068].

\bibitem{Doran:2002ec}
M.~Doran and C.~Wetterich,
Nucl.\ Phys.\ Proc.\ Suppl.\ {\bf 124}, 57 (2003)
[arXiv:astro-ph/0205267].

\bibitem{Townsend:2001ea}
P.~K.~Townsend,
JHEP {\bf 0111}, 042 (2001)
[arXiv:hep-th/0110072].

\bibitem{Andrianopoli:1996cm}
L.~Andrianopoli, M.~Bertolini, A.~Ceresole, R.~D'Auria, S.~Ferrara, P.~Fr\'e, 
and T.~Magri,
J.\ Geom.\ Phys.\ {\bf 23}, 111 (1997)
[arXiv:hep-th/9605032].

\end{thebibliography}
\end{document}